\documentclass[conference]{IEEEtran}
\IEEEoverridecommandlockouts
\usepackage{cite}
\usepackage{amsmath,amssymb,amsfonts}
\usepackage{algorithmic}
\usepackage{graphicx}
\usepackage{textcomp}
\usepackage{xcolor}
\usepackage{booktabs}
\usepackage{adjustbox}
\usepackage{setspace}

\def\BibTeX{{\rm B\kern-.05em{\sc i\kern-.025em b}\kern-.08em
    T\kern-.1667em\lower.7ex\hbox{E}\kern-.125emX}}
\begin{document}

\title{Hardware-level Interfaces for Hybrid Quantum-Classical Computing Systems}

\author{\IEEEauthorblockN{Konstantinos Rallis\IEEEauthorrefmark{1}\IEEEauthorrefmark{2}, Ioannis Liliopoulos\IEEEauthorrefmark{1}\IEEEauthorrefmark{2}, Evangelos Tsipas\IEEEauthorrefmark{1}\IEEEauthorrefmark{2}, Georgios D. Varsamis\IEEEauthorrefmark{1}\IEEEauthorrefmark{2}, Nikolaos Melissourgos\IEEEauthorrefmark{3},} \IEEEauthorblockN{Ioannis G. Karafyllidis\IEEEauthorrefmark{2}, Georgios Ch. Sirakoulis\IEEEauthorrefmark{2}, Panagiotis Dimitrakis\IEEEauthorrefmark{1}}

\IEEEauthorblockA{\IEEEauthorrefmark{1}\textit{National Centre of Scientific Research "Demokritos"}, \IEEEauthorrefmark{2}\textit{Democritus University of Thrace},}  
\IEEEauthorblockA{\IEEEauthorrefmark{3}\textit{Hewlett Packard Enterprise Greece}}
}

\maketitle

\begin{abstract}
The technology of Quantum Computing (QC) is continuously evolving, as researchers explore new technologies and the public gains access to quantum computers with an increasing number of qubits. In addition, the research community and industry are increasingly interested in the potential use, application, and contribution of QCs to large-scale problems in the real world as a result of this technological enhancement. QCs operations are based on quantum mechanics, and their special properties are mainly exploited to solve computationally intensive problems in polynomial time, problems that are commonly unsolvable, even by High-Performance Computing systems (HPCs) in a feasible time. However, since QCs cannot perform as general-purpose computing machines, alternative computational approaches aiming to boost further their enormous computing abilities are requested, and their combination as an additional computing resource to HPC systems is considered as one of the most promising ones. In the proposed hybrid HPCs, the Quantum Processing Units (QPUs), similar to GPUs and CPUs, target specific problems that, through Quantum Algorithms, can exploit quantum properties like quantum entanglement and superposition to achieve substantial performance gains from the HPC point of view. This interconnection between classical HPC systems and QCs towards the creation of Hybrid Quantum-Classical computing systems is neither straightforward nor standardized while crucial for unlocking the real potential of QCs and achieving real performance improvements. The interconnection between the classical and quantum systems can be performed in the hardware, software (system), or application layer. In this study, a concise overview of the existing architectures for the interconnection interface between HPCs and QCs is provided, focusing mainly on hardware approaches that enable effective hybrid quantum-classical operation.
\end{abstract}


{\setstretch{0.93}
\section{Introduction}
\label{intro}

Continuous advances in computing technology have made it an integral part of daily life, extending its field of application both across industry and research sectors. Especially the rise of the data era and the widespread adoption of artificial intelligence in almost every industry and research branch for the solution of computationally intensive problems have significantly increased the need for more powerful computing resources. This new and ever-evolving condition has sparked significant interest in high-performance computing (HPC) systems, which have become tightly connected to several industrial applications and research. HPC systems are continuously developing and improving through optimization and innovation, with efforts driven by both public and private initiatives around the world. New systems are being installed and provided as a service or in other forms by several different vendors \cite{milojicic2021future}.

The aforementioned systems are based on conventional computing nodes, which are either CPUs, GPUs, or any other similar co-processor or application-specific accelerator. In parallel to conventional computing and HPC systems, Quantum Computing (QC) has emerged as a completely novel and revolutionary computing paradigm that operates based on quantum mechanical phenomena \cite{rietsche2022quantum}. This new type of computing promises to significantly accelerate the solution of complex problems which many times are beyond the reach of classical computing and unsolvable in a reasonable timeframe. While quantum computers promise to provide solutions to NP-hard and NP-complete problems in polynomial time, they are not yet capable of operating as general-purpose computing systems. Instead, they target specific problem categories that can significantly benefit from their operation principles \cite{ruefenacht2022bringing}, while for doing this, they also physically resemble HPC infrastructures that require specialized installation and maintenance, as well as specific operational frameworks. In its current state, QC technology is being developed mainly by international technology firms that provide access to quantum resources, in most cases via the cloud, with continuous improvements in terms of qubit scalability, availability, and computational resilience \cite{de2021materials, memon2024quantum, kim2023evidence}.

Unlocking the full potential of QCs and Quantum Processing Units (QPUs) is not straightforward, as the effective execution of quantum programs requires specialized problem mapping, input and output data handling, job scheduling and resource management, tasks that still cannot be performed by the quantum system alone. Several types of problems are suitable for execution and solution in QCs, but yet there is no universal automated process that seamlessly translates classical problems into quantum-solvable formats, as this is an area under active research \cite{bar2025approach}. Currently, the most viable approach to utilizing quantum computing involves its co-existence and seamless integration with classical HPC systems. In this approach, the Quantum Processing Units (QPUs) function as different types of additional computational resources, similar to the CPU and GPU nodes of existing HPC systems. In this hybrid setup that combines classical and quantum resources (one or more QPUs), the HPC system, in addition to dealing with computing tasks, is also responsible for algorithm transpiling, orchestration of task execution, job scheduling, data preparation, and communicating with QPUs \cite{elsharkawy2024integration}.

The effective integration of QPUs into HPC infrastructures is a substantial step toward unlocking the complete potential of quantum computing, which is in turn a driving force behind the development of QC technology in total. This integration can be performed at both software and hardware level \cite{beck2024integrating} and this work provides a short but comprehensive overview of existing HPC-QC integration strategies, with a particular emphasis on hardware level interfacing, which will, to a large extent, also affect the software-level integration. In Section \ref{hybrid}, a detailed presentation of the possible advantages of the combination of HPC and QC systems is provided in accordance with the presentation of the two (2) main integration models. Then, in Section \ref{hardware_interface}, the four (4) main integration architectures are thoroughly analyzed, offering an overview of the basic physical means that are employed for HPC-QC integration, together with an evaluation of their advantages and limitations. Finally, Section \ref{conclusion} concludes the paper with a short reference to the existing open issues of HPC-QC hardware-level integration.

\section{Hybrid Quantum Classical Computing Systems}
\label{hybrid}

As it has been mentioned, classical HPC systems struggle with providing soultions to NP-hard and NP-complete problems. The solution time of such problems scales exponentially with the size of the problem and thus requires significant computational resources. The execution times grow abruptly beyond practical limits with the increase of the dimensions of the inputs. There are many problems that lay in the categories of NP-hard/NP-complete, which are commonly encountered in research and industrial applications \cite{panchal2015solving}. Quantum Computing, by taking advantage of quantum phenomena such as superposition and entanglement, promises to allow better handling and faster solution of such problems, offering exponential speedups \cite{chatterjee2024solving}. Representative types of problems that can benefit from the use of quantum computing are several combinatorial optimization problems \cite{gemeinhardt2023quantum}, machine learning and AI-related tasks \cite{liliopoulos2025hybrid}, cryptography and security tasks \cite{radanliev2024artificial}, computational biology tasks \cite{varsamis2023quantum}, and computational chemistry tasks \cite{mcardle2020quantum} just to name a few.
However, as QPUs are not general-purpose processing units and cannot operate as standalone systems, at the current technology level, they usually rely on the interaction with classical general-purpose computers, which will provide support related to data preparation, algorithm mapping, and orchestration, as well as output data post-processing. This is one of the most significant reasons that significantly increases the need and possible impact of the interfacing of HPCs with QPUs towards the creation of hybrid Quantum-Classical computing environments.

Additionally, there are already several well-known quantum algorithms, like the Variational Quantum Eigensolver (VQE) \cite{tilly2022variational}, the Quantum Approximate Optimization Algorithm (QAOA) \cite{blekos2024review}, and many more, that can be considered hybrid by nature, as they do not require the use of HPC just for pre and post-processing of data and inputs. They are based on iterations, requiring a tight interplay between classical and quantum computation in order to provide a solution efficiently and accurately \cite{baaquie2023quantum}. Thus, the total performance and efficiency of such kind of quantum-classical workflows heavily depend on the quality of the interfacing between HPC and quantum systems.

As of today, the HPC-QC hardware interfacing techniques span from loosely integrated to tightly integrated system approaches, with the most realistic concept being the loose one, at the current Technology Readiness Level (TRL) of QPUs and QCs in general.
The \textit{loose} integration approaches involve quantum devices, which are being accessed remotely, via some kind of local communication network, or the cloud. On the other hand, the \textit{tight} integration approaches involve direct access of the QPUs by the HPC, meaning that QPUs and HPC resources are connected physically through a common node, similar to other classical computing resources (CPUs, GPUs) \cite{beck2024integrating}. While software-level interfacing keeps evolving, taking heavy inspiration from HPC systems, especially in job managing and scheduling, hardware interfacing proves to be a bottleneck, entailing a set of challenges that need to be overcome in order to achieve low-latency and high-bandwidth interconnection between quantum and classical resources and unlock the full potential of QC as an accelerator. 

In the following Section, the existing hardware-layer interfacing techniques will be examined and compared in terms of architecture and physical means, which enable efficient interfacing and seamless interaction between HPC systems and QPUs.

\begin{figure*}[t!]
    \centering
    \includegraphics[width=0.80\textwidth]{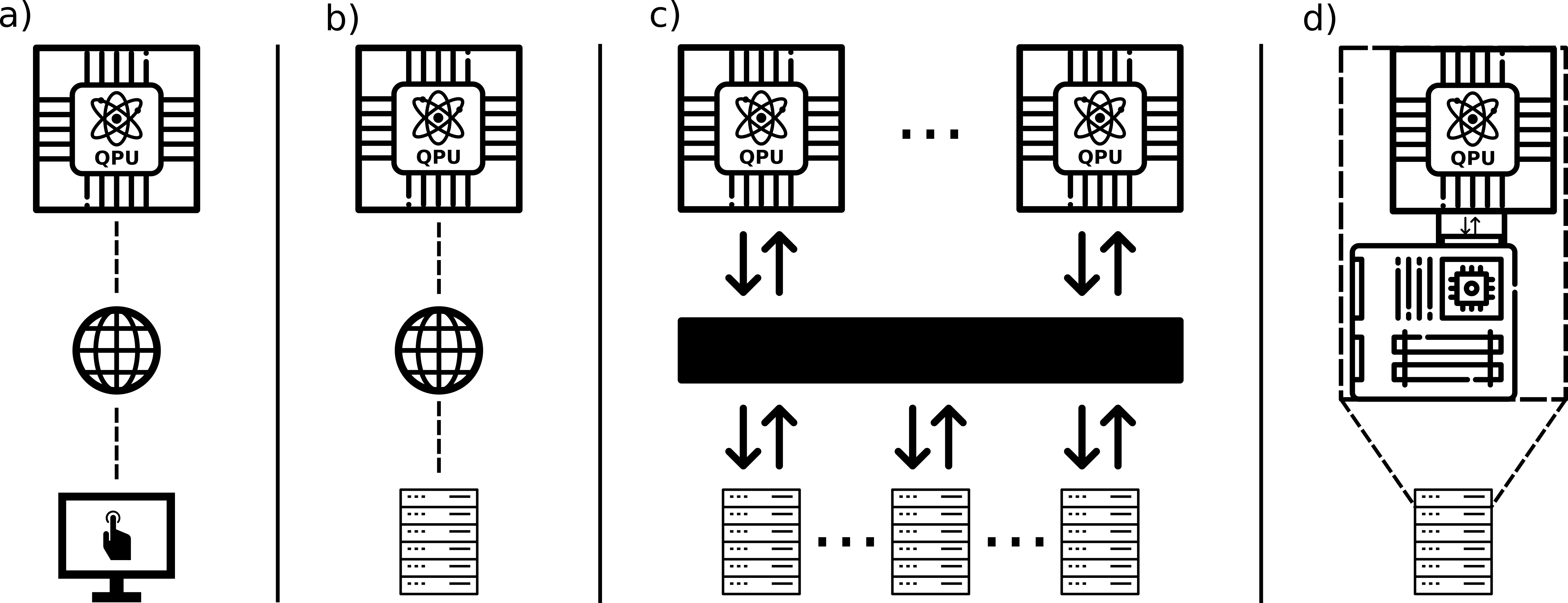}
    \caption{a) Standalone QPU - Loose integration: User interacts with a single QC through a web interface. b) Co-location - Loose Integration: The HPC interacts with a single QPU. They are two distinct infrastructures which are physically near and communicate through common local network or even apart from each other and communicate via cloud services c) Co-location - Tight Integration: The HPC system interacts with multiple QPUs. They are physically near but still separate hardware infrastructures which communicate via common hardware high-speed interconnects d) On-node - Tight Integration: The QPU is now embedded inside the classical computing node in the form of an external co-processor like a GPU.}
    \label{fig1}
\end{figure*}

\section{Hardware Layer HPC QC Interfacing Approaches}
\label{hardware_interface}
\subsection{Architectural Approaches}
\label{architectural}
Towards the creation of Hybrid HPCQC systems, the approach that will be followed on the hardware-level interfacing of the corresponding components is crucial, as it will affect several aspects of the whole system, its total complexity and performance, and will also define the requirements and the needed components of the software stack for software-level interfacing.

The interfacing between HPC and QC systems can be clustered, among others, into three (3) main distinct categories, based on the followed microarchitecture and thus the relative location between the two physical subsystems and their interaction. Those three ($3$) categories are: 1) Standalone, 2) Co-located, and 3) On-node integration \cite{elsharkawy2023integration}.

\textbf{1) Standalone:} The $1^{st}$ category, refers to the case where the QPU operates as an isolated system and is not in any automated way interconnected with an HPC system, following the principles of the \textbf{loose integration model}. A conventional computing system is always present in order to allow the user to provide the required information to the QPU, most commonly via a web interface, as seen in Fig.~\ref{fig1}(a). However, in this case, the user has to manually prepare the problem by defining it and selecting or developing the required algorithm \cite{rohe2024problem}. Then, the developed quantum algorithm has to be mapped to a quantum-ready high-level representation form, mainly in the form of a Quantum Circuit, through high-level programming languages. Following, the quantum circuit optionally undergoes a set of specific optimizations before being transpiled to hardware-specific instructions and executed on a specifically selected QPU, providing back the results for post-processing \cite{javadi2024quantum}. The standalone operation does not support any type of HPC-QC combined workflows, providing minimal integration. It is, however, suitable for the development, testing, and evaluation of Quantum Algorithms, as well as for the experimentation with specific QPUs and qubit technologies for further investigating their technology-related properties and behavior \cite{weder2020integrating}.

\textbf{2) Co-location:} The $2^{nd}$ category refers to systems that entail co-location of HPC and QC hardware. Co-location practically means that HPC and QC resources are physically near, communicating via a common network, or even isolated and communicating via the cloud. In that type of interfacing, two ($2$) main different setups exist, including either co-location in a networking manner or physical co-location.

The first setup, the co-location through a common communication network, lays on the principles of the \textbf{loose integration model}. Now the HPC system, with the available classical computing nodes, can establish communication with a single QPU as seen in Fig.~\ref{fig1}(b), to complete also hybrid quantum-classical workloads. The flow that is commonly followed by such systems is similar to the one described above, beginning with the preparation of the quantum algorithm and concluding with the return of the produced results for post-processing. One of the key differences is that some steps, which required manual execution in the standalone case, are now automated \cite{humble2021quantum}. Also, this kind of interconnection allows the execution of hybrid iterative algorithms, as mentioned in Section \ref{hybrid}, that require repetitive data exchange between classical and quantum nodes. However, the efficient accomplishment of this automated communication requires a more advanced, complicated, and sophisticated toolset for managing the whole system. Due to the complexity of installing and managing QPUs, the employment of connections over the cloud, instead of a local communication network is also common, as many QPUs are offered as a service (QaaS) \cite{ahmad2024reference}. This loose co-location also has to overcome the latency introduced due to the network-based communication, which can become a bottleneck and penalize the efficiency and overall performance of the whole system. If the communication networks are publicly available and especially in the case of cloud access, the problem of data security is also raised \cite{golec2024quantum}.

The second setup includes physical co-existence of the resources and follows the \textbf{tight model of integration}. It includes the physical co-existence of more than one QPU, interconnected and communicating with one another as well as with classical computing nodes, as seen in Fig.~\ref{fig1}(c). Here, multiple QPUs, even from different vendors, and classical resources remain in separate but closely connected units using specialized low-latency hardware interfaces. That way, more quantum computing power can become available and directly accessible, allowing optimized data sharing and more intense collaboration between classical and quantum computations and offering reduced latency. The higher amount of QPUs allows for better orchestration of multiple tasks and enables the handling of more complex quantum algorithms, which are also able to be executed on multiple QPUs, in a way similar to classical parallel computing \cite{pastor2024circuit}. As a more complex scheme of integration, this setup requires a more sophisticated software stack for its seamless operation, to be able to handle the heterogeneity in terms of the QPU technology and efficiently cover the demands for scheduling and data exchange.

\textbf{3) On-node:} This $3^{rd}$ category refers to the on-node integration of QPUs directly into computing infrastructures, inside the HPC node as seen in Fig.~\ref{fig1}d), in accordance with the integration of other accelerators, like GPUs or TPUs, following the principles of the \textbf{tight integration model}. This is the ultimate goal of the realization of Hybrid HPC-QC systems, and its implementation spans from a direct connection of the QPU on a motherboard, to a more complicated and futuristic integration of classical resources and QPUs using chiplets, to form an undivided quantum-enhanced system \cite{ruefenacht2022bringing}. This approach leads to the QPU to operate in combination with other classical computing resources, down to the instruction level, enabling the execution of hybrid algorithms, where the classical and quantum components of the system automatically interact. As a result, QPUs are converted into efficient application-specific accelerators for specific problem categories that can benefit the most from their specialized capabilities, as mentioned above \cite{chen2024multi}. Such a physical co-location allows for real-time quantum-classical operations, overcoming the common problem of increased latency in the communication between classical and quantum resources through the network in other interfacing approaches. Adding to that, it also guarantees data security, especially compared with integration methodologies that share data over public networks. It is, however, the hardest to achieve, as it requires extreme hardware and software complexity in order to achieve smooth communication and orchestration of the available resources. Additionally, the challenge, especially for the integration of QPUs on the same package with classical resources, is compounded by the highly intricate fabrication technologies involved in developing qubits, as well as their operation in ultra-low temperature environments.

A summarized version of those four (4) different setups that lay in the three (3) described categories, can be viewed in Tab.~\ref{tab:hpc_qc_integration_comparison}.

\begin{table*}[htbp]
\centering
\caption{Comparison of HPC–Quantum Integration Architectures}
\label{tab:hpc_qc_integration_comparison}
\renewcommand{\arraystretch}{1.1}
\begin{adjustbox}{width=\textwidth}
{\small 
\begin{tabular}{|p{3.2cm}|p{3.2cm}|p{3.2cm}|p{3.2cm}|p{3.2cm}|}
\hline
\textbf{Integration Type} & \textbf{Advantages} & \textbf{Limitations} & \textbf{Typical Interconnects} & \textbf{Use Cases} \\
\hline
\textbf{Standalone Integration} (Loose – Standalone) &
Simple setup. Easy access through cloud providers. Good for development and education. &
Very high latency. Manual workflows. No real-time hybrid execution. &
WAN / Internet, Web APIs. &
Quantum algorithm prototyping, educational purposes. \\
\hline
\textbf{Loose Co-Located Integration} (Loose – Co-located) &
Lower latency vs. Standalone. Local control of QPU. Basic hybrid workflows are possible. &
Network latency persists. Requires on-site QC hardware. Separate resource management. &
Ethernet, InfiniBand, Internet. &
Hybrid VQE/QAOA algorithms, on-premises hybrid experiments. \\
\hline
\textbf{Tight Co-Located Integration} (Tight – Co-located) & 
Low latency. Supports multi-QPU setups. Unified resource scheduling. &
Complex integration. Heterogeneous QPU vendor issues. Advanced orchestration is required. &
PCIe Gen4/5, CXL, InfiniBand, Experimental Quantum Networking. &
Quantum chemistry, combinatorial optimization, multi-QPU hybrid HPC. \\
\hline
\textbf{On-Node Integration} (Tight – On-node) &
Near-zero latency. Real-time classical-quantum interaction. Enables adaptive hybrid workloads. &
Extreme hardware complexity. Cryogenic challenges. Still in research stage, no products yet. &
Direct PCIe, CXL, Cryo-CMOS controllers, Chiplet integration (future). &
Real-time hybrid HPC-QC, adaptive quantum error correction, next-gen HPC accelerators. \\
\hline

\end{tabular}
} 
\end{adjustbox}
\end{table*}

\subsection{Physical interconnects between QPUs and HPC systems}

\textbf{Classical Peripheral Interfaces:} It is common for the realization of Hybrid Quantum Classical computing systems, to treat the quantum part, the QPU and mainly its control unit, as an additional computing resource, as a peripheral device exactly like a GPU. Based on this principle, the interconnection between HPC and QPU can be implemented with common standard buses like PCI Express (PCIe). A distinctive example is the use of PCI Express Gen 5 by the NVIDIA DGX Quantum system \cite{nvidiaDGXQuantum}. Extending this, another type of connection that can be exploited for the interfacing of QPUs with HPCs is the Common Express Link (CXL), which is built upon the physical and electrical interface of the PCIe. Both provide high-bandwidth and low-latency but CXL link offers additional features like enabling cache-coherence and allowing for memory sharing \cite{beck2024integrating}. Additionally, all these interfaces commonly incorporate FPGA-based controllers, which are leveraged to generate the required quantum control pulses and readout signals, providing precise timing control and fine-grained, flexible, and reprogrammable interaction between classical and quantum components \cite{xu2021qubic}. Seeking increased performance compared to FPGAs, especially in terms of efficiency and integration density, ASIC-based quantum controllers can also be used.

\textbf{High-Speed Networks (InfiniBand, Ethernet):}
As previously mentioned, the most common case based on the current technology level, is to have Hybrid HPC QC systems that follow the loose integration model. For such models, where most of the time the QC resources are at a remote location, in a separate cryostat infrastructure, it is common to establish communication with the HPC resources over the network. This happens mainly with the use of InfiniBand or high-speed Ethernet infrastructures, which are used for the exchange of job data and results between the HPC host and QPU's control unit \cite{beck2024integrating}. The first one, InfiniBand, enables Remote Direct Memory Access (RDMA) and offers ultra-low latency and high bandwidth \cite{shpigelman2022nvidia}. This is the case for quantum systems, that do not communicate through a common physical node but instead via a local network.

\textbf{Cryogenic Wiring and Microwave Links:}
There are several different technologies for the fabrication of qubits, the basic unit cell of quantum computing. However, as the operation of qubits requires very low noise levels, most of the existing technologies require very low temperatures in the scale of a few millikelvins. That practically means that in most cases, the qubits need to be at the cryostat, as well as their control electronics. 

For this reason, several cryo-CMOS based controller architectures have been proposed and investigated for driving the qubits. They are electronics that can be placed and operate near the qubits, at cryogenic environments, providing the required control signal generation, possible signal amplification when required during the readout, some first-layer local signal processing. Their use targets significantly simplifying physical connections, improving performance, reducing latency and noise, and finally contributing to the scalability of quantum systems \cite{staszewski2021cryo,oka2022cryo, hornibrook2015cryogenic, sebastiano2020cryo}.

Thus, for the interconnection of the HPC electronics, which are at room temperature, with the qubits and their control electronics, which are at the cryostat, two (2) possible different physical means can be employed: coaxial cables, which are flexible and support both DC biases and microwave signals, and waveguides, which are rigid structures designed specifically for very high-frequency microwave signals with lower attenuation and reduced thermal conduction. Either of those two in combination with control lines is responsible for the transmission of microwave pulses, DC biases, and readout signals and is practically the physical means to interconnect PCIe or FPGA-based controllers of the HPC part, with the QPU. For just a single qubit, commonly two input lines are required for its control and multiple I/O lines for the readout \cite{das2024chip}. In order to counter this problem that will get more severe with the scale-up of QPUs, several techniques have already been proposed, like the use of cryogenic RF switches and crossbar designs \cite{potovcnik2021millikelvin, ciobanu2024optimal, li2018crossbar}. Additionally, this type of physical means can employ techniques like frequency multiplexing, in order to increase channel density. There are several techniques under investigation, like cyro-electronic multiplexers \cite{acharya2023multiplexed} or photonic fiber links with additional techniques like wavelength division multiplex \cite{joshi2023scaling}, which could further increase density by replacing several coaxial cables with only one fiber that will carry the signals to the cryo part under specific optical modulation \cite{lecocq2021control}. 

\textbf{Quantum Networking:}
 As mentioned previously, in Section \ref{architectural}, hybrid systems that entail more than one (1) QPU even from multiple vendors can exist. In order to achieve tighter integration and to be able to combine the provided resources from the available QPUs for the solution of more complex problems that require a bigger amount of qubits, a method for direct communication between the QPUs is needed. This communication between QPUs via quantum entanglement links without the intervention of classical HPC resources is called Quantum Networking. Different types of interconnections can be realized for the transfer of quantum states between qubits and QPUs, based on the type of integration and the technology of the qubits \cite{escofet2023interconnect}. Those emerging interconnection techniques involve the use of waveguides or optical fibers for photonic-based communication over long distances \cite{marinelli2023dynamically}, inter-chip interconnects that directly connect qubits on different chips that are physically near and vary based on the qubit technology, including among others, coaxial cables, superconducting transmission lines and capacitive coupling via resonators \cite{escofet2023interconnect}, quantum teleportation that exploits correlated photon pairs to transfer quantum states between qubits remotely \cite{hu2023progress}, ion shuttling that finds application in ion-trapped based qubits and involves the physical relocation of qubits in close proximity \cite{schoenberger2024shuttling}, and others. All those different types of quantum links are still in development and experimentation, paving the way for the future of quantum computing, where there will be the transmission of qubits instead of bits, enabling distributed or parallel quantum computing. The use of those quantum entanglement links can ultimately lead to the creation of quantum local networks, with multiple entangled QPUs that will be managed by HPC resources \cite{azuma2023quantum, barral2024review}.

The use of those physical interconnects in HPC-QC integrated systems of different architectures, can be viewed in Tab.~\ref{tab:hpc_qc_integration_comparison}.

\section{Conclusions}
\label{conclusion}
In summary, this work provides a comprehensive overview of the existing hardware interfaces for merging High-Performance Computing (HPC) systems with Quantum Computers and their Quantum Processing Units (QPUs), towards the creation of hybrid quantum-classical computing systems. The main different architectural approaches on the interfacing between HPC and QC subsystems have been presented, outlining their advantages and limitations in terms of complexity, latency, and overall performance. As an emerging domain under research, it still entails a set of open challenges and significant unresolved issues. One of them is the trade-off between ultra-low latency and reliability of communication. Adding to that, the lack of standardization in HPC-QC interfacing leads to fragmentation through several custom solutions. Finally, scalability is one of the biggest issues and will become even more prominent with the increase of the provided amount of qubits and QPUs, which will make the task of their interfacing and managing the data traffic while also maintaining low error levels, even more difficult. All those open issues ensure that the uninterrupted development and investigation of novel and existing hardware interfaces will determine the efficiency, scalability, and usability of future hybrid quantum-classical computing systems, and significantly influence their future advancement.

\section*{Acknowledgment}
This research has been supported by the project “A catalyst for EuropeaN ClOUd Services in the era of data spaces, high-performance and edge computing(NOUS)”, Grant Agreement Number 101135927. Funded by the European Union, views and opinions expressed are, however, those of the authors only and do not necessarily reflect those of the European Union. Neither the European Union nor the granting authority can be held responsible for them.

}
\bibliographystyle{IEEEtran}
{\setstretch{0.97}
\bibliography{biblio}
}

\end{document}